\documentclass[%
 reprint,
superscriptaddress,
nofootinbib,
 amsmath,amssymb,
pra,
longbibliography
]{revtex4-2}
\usepackage{dcolumn}
\usepackage{bm}
\usepackage{graphicx}
\usepackage{amsthm}
\usepackage{color}
\usepackage{xcolor}
\usepackage{verbatim}
\usepackage{esint}
\usepackage[english]{babel}
\usepackage{float}
\usepackage{amsfonts}
\usepackage{slashed}
\usepackage{latexsym}
\usepackage{bm}
\usepackage[colorlinks = true,
            linkcolor = blue,
            urlcolor  = blue,
            citecolor = blue,
            anchorcolor = red]{hyperref}
\usepackage{esint}
\usepackage{soul}
\usepackage{cancel}
\usepackage[normalem]{ulem}
\usepackage{empheq}
\usepackage{dsfont}
\usepackage{amsmath}
\usepackage{ulem}
\usepackage{epsfig}
\usepackage{tikz}

\newcommand\be{\begin{equation}}
\newcommand\ee{\end{equation}}
\newcommand\la{\langle}
\newcommand{\ra}{\rangle}

\begin{document}
\title{Beyond superoscillation: General theory of approximation with bandlimited functions}

\author{Tathagata Karmakar}
\email{tkarmaka@ur.rochester.edu}
	\affiliation{Department of Physics and Astronomy, University of Rochester, Rochester, NY 14627, USA}
	\affiliation{Center for Coherence and Quantum Optics, University of Rochester, Rochester, NY 14627, USA}
\affiliation{Institute for Quantum Studies, Chapman University, Orange, CA 92866, USA}
\author{Andrew N. Jordan}
\email{jordan@chapman.edu}
\affiliation{Institute for Quantum Studies, Chapman University, Orange, CA 92866, USA}
\affiliation{Department of Physics and Astronomy, University of Rochester, Rochester, NY 14627, USA}
	\affiliation{Center for Coherence and Quantum Optics, University of Rochester, Rochester, NY 14627, USA}

\date{\today}

\begin{abstract}
We give a general strategy to construct superoscillating/growing functions using an orthogonal polynomial expansion of a bandlimited function. The degree of superoscillation/growth is controlled by an anomalous expectation value of a pseudodistribution that exceeds the band limit. The function is specified via the rest of its cumulants of the pseudodistribution. We give an explicit construction using Legendre polynomials in the Fourier space, which leads to an expansion in terms of spherical Bessel functions in the real space. The other expansion coefficients may be chosen to optimize other desirable features, such as the range of super behavior. We provide a prescription to generate bandlimited functions that mimic an arbitrary behavior in a finite interval. As target behaviors, we give examples of a superoscillating function, a supergrowing function, and even a discontinuous step function. We also look at the energy content in a superoscillating/supergrowing region and provide a bound that depends on the minimum value of the logarithmic derivative in that interval. Our work offers a new approach to analyzing superoscillations/supergrowth and is relevant to the optical field spot generation endeavors for far-field superresolution imaging.
   
\end{abstract}

\maketitle

Superoscillation occurs when a function oscillates faster in a region than the band limits of the function \cite{berry2019roadmap}.  Similarly, supergrowth refers to a local growth or decay rate that exceeds the band limit \cite{jordan2020superresolution}. Both phenomena correspond to changes on a scale smaller than the smallest wavelength present in a function's Fourier decomposition. Therefore, they can both help  resolve subwavelength objects in diffraction-limited systems. As supergrowing regions usually bridge superoscillatory spots with high-intensity side lobes, the former regions are expected to contain significantly higher light than the latter ones \cite{SG_imaging_drafft}. Supergrowth is a recently introduced concept. However,  superoscillatory spots are routinely utilized in unlabeled far-field superresolution imaging \cite{chen_superoscillation_2019}. Experimental and numerical investigations to generate and optimize superoscillatory spots have been performed \cite{baumgartl_far_2011,kozawa_numerical_2015,Kozawa2018Feb,diao_controllable_2016, rogersOptimisingSuperoscillatorySpots2018,hu_optical_2021,36243494}.

While numerous examples are known, general recipes for constructing such functions remain elusive.  We mention the work of Achim Kempf and collaborators \cite{kempfBlackHolesBandwidths2000,kempf2018four}, who have both additive \cite{ferreira2007construction} and multiplicative strategies \cite{chojnacki2016new,soda_efficient_2020}, involving the progressive constructions of a superoscillatory function using basic superoscillating functions as building blocks.  Other methods have been discovered to cause the function to superoscillate everywhere in a suitable mathematical limit \cite{aharonov2021new}. Additionally,  the stability of superoscillatory properties in the presence of perturbance in generating coefficients has been studied \cite{tangScalingPropertiesSuperoscillations2016}. 

\begin{figure}[tbh]
    \centering    \includegraphics[width=\linewidth]{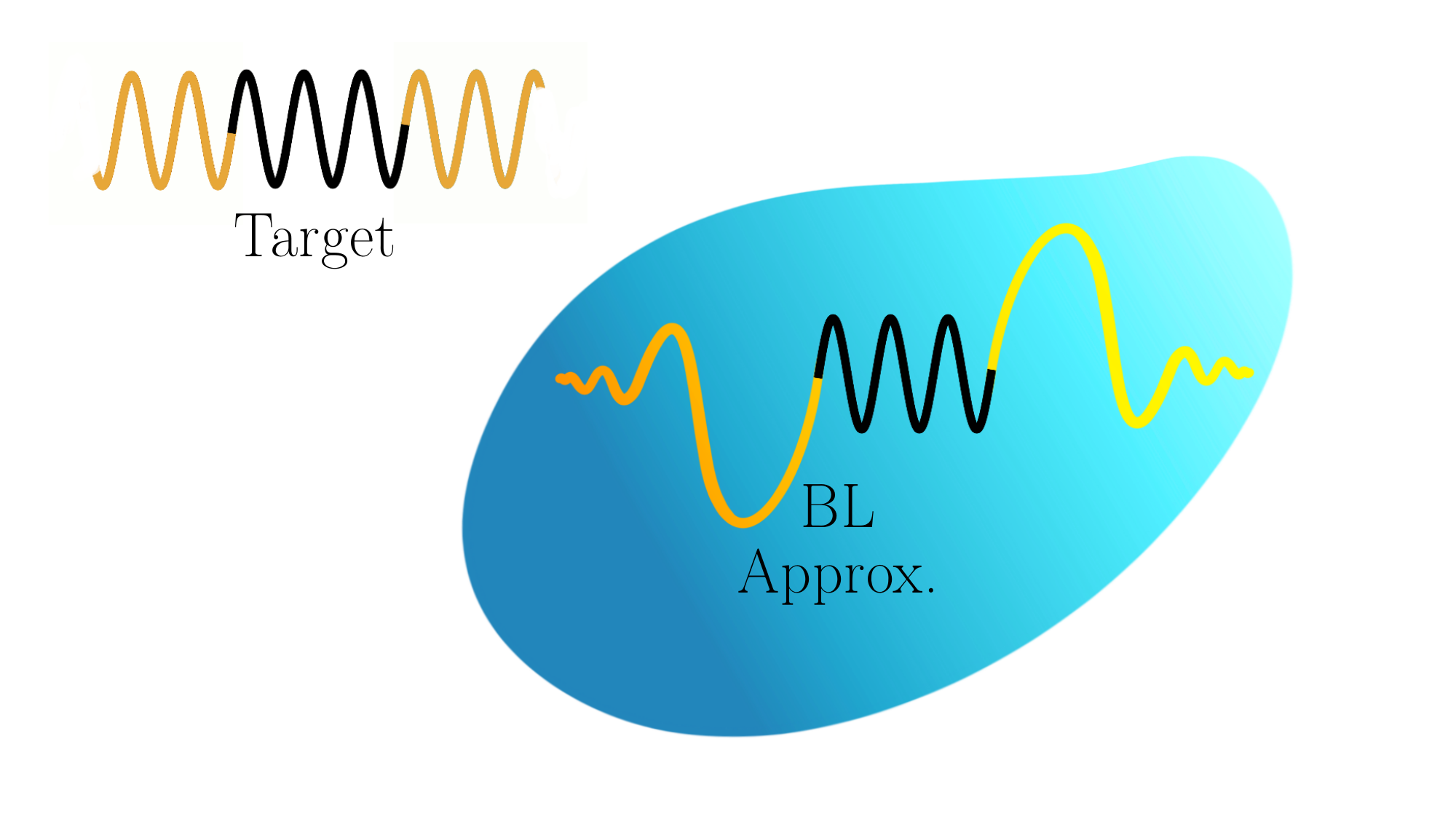}    \caption{In this  schematic, the shaded region denotes the space of bandlimited functions. We prescribe a method to find the bandlimited approximation of a target function (existing outside the bandlimited space of functions) in an arbitrary finite interval.}
    \label{fig:schematic}
\end{figure}

In this work, we build a robust framework suitable for analyzing and controlling the local behavior of a bandlimited function.  First, we demonstrate that a function's local wavenumber/growthrate can be expressed in terms of the cumulants of its Fourier transform. This constitutes a parallel account of the function and its superoscillatory/supergrowing properties in terms of a pseudodistribution. Next, by adopting the Legendre polynomials as a basis in the Fourier space (consequently, spherical Bessel functions in the original space), we gain direct access to the amount of superoscillation/supergrowth at the origin through the corresponding coefficients. We then present a method to approximate a continuous function in an interval in terms of spherical Bessel functions. Such a scheme lets us create bandlimited functions that mimic any desirable superoscillating/supergrowing behavior in arbitrary finite intervals, as the Fig.~\ref{fig:schematic} depicts below.
On a related note, Chremmos et al.~\cite{chremmos_superoscillations_2015}  proposed a methodology to approximate polynomials by multiplying them with an envelope function from Paley-Wiener space. Also, Šoda et al.~\cite{soda_efficient_2020} showed using MacLaurin series expansion, smooth functions can be approximated in terms of bandlimited functions with arbitrary accuracy. However, the prescription provided in our analysis does not require any polynomial approximation or derivative calculations of the target function.
Moreover, as we will see, the method applies to any general continuous function and can work well with discontinuous functions too. We also provide a bound on the fractional energy available in a function's superoscillating/supergrowing region in terms of its logarithmic derivative. This complements the work of Ferreira et al.~\cite{ferreiraSuperoscillationsFasterNyquist2006}, who found that the energy required to create a superoscillatory signal increases exponentially with the number of oscillations. Our work explores and provides a natural choice of basis functions for functions defined on the real line. The investigations here give a very useful and flexible device for numerical optimization and superoscillatory/supergrowing optical field spot generation.

This paper is organized as follows. In Sec.~\ref{gen}, we look at a function in Fourier space and describe its local derivatives in terms of cumulants in Fourier space. Sec.~\ref{legendre} describes a bandlimited function in terms of Legendre polynomials in its Fourier space and Sec.~\ref{real_space} in terms of spherical Bessel functions in real space. In Sec.~\ref{moments}, we calculate the higher-order moments of the function in Fourier space and provide its power series expansion.  We present our approximation scheme in Sec.~\ref{approx_scheme}, where Secs.~\ref{approx_real} and ~\ref{approx_cyl} describe the method for functions on the real line and cylindrically symmetric functions, respectively. Sec.~\ref{enerbound} provides a bound on the fractional energy in a supergrowing/superoscillating region.  Finally, we conclude in Sec.~\ref{discussion}.  

\section{\label{gen}General Principles}
Here we take a first principles approach, and consider a band-limited function ${\tilde f}(\omega)$ in the Fourier space \footnote{Here we take the convention that $f(t)=\int_{\Omega_{min}}^{\Omega_{max}}\tfrac{d\omega}{\sqrt{2\pi}}e^{i\omega t}\tilde{f}(\omega)$, such that the inverse Fourier transform is $\tilde{f}(\omega)=\int_{-\infty}^{\infty}\tfrac{dt}{\sqrt{2\pi}}e^{-i\omega t}f(t)$.}, that has non-zero value only between $\Omega_{min}$ and $\Omega_{max}$.  We first redefine the Fourier variable to 
\be
\omega' = \frac{ 2\omega - \Omega_{max} - \Omega_{min}}{\Omega_{max} - \Omega_{min}}.
\ee
The original function $f$ can be expressed in terms of dimensionless frequency and time as:
\be
f(\tfrac{2}{\Delta\Omega}t') = \exp\left[i \frac{2\bar \Omega}{\Delta \Omega} t'\right] \int_{-1}^1 \tfrac{d\omega'}{\sqrt{2\pi}} e^{i \omega^\prime t'} \frac{\Delta \Omega}{2} {\tilde f}(\tfrac{\Delta\Omega}{2}\omega'+\bar{\Omega}),
\ee
where $\Delta \Omega = \Omega_{max} - \Omega_{min}$, and ${\bar  \Omega} = (\Omega_{max} + \Omega_{min})/2$, and $t' = \Delta \Omega\, t/2$. We can redefine $g(t')=\tfrac{2}{\Delta\Omega}f(\tfrac{2}{\Delta\Omega}t')\exp\left[-i\frac{2\bar{\Omega}}{\Delta\Omega}t'\right]$ and $\tilde{g}(\omega')=\tilde{f}(\tfrac{\Delta\Omega}{2}\omega'+\bar{\Omega})$. Thus for any bandlimited function, we have the dimensionless Fourier expansion
\be
g(t')=\int_{-1}^{1}\tfrac{d\omega'}{\sqrt{2\pi}}e^{i\omega' t'}\tilde{g}(\omega').
\ee
Therefore, we focus on the dimensionless frequency defined on the interval $[-1, 1]$ and drop the primes for notational convenience. Throughout the rest of our analysis, the term `bandlimited' refers to functions with only non-zero Fourier spectrum  in the $[-1,1]$ interval. 

We note that the Fourier transform of a bandlimited function
$g(t)$ can be viewed as a generating function of the cumulants of a pseudodistribution $ {\tilde g}(\omega)$,
\be
i^{-n} \frac{d^n}{dt^n} \ln g(t)\vert_{t=0} = \langle\!\langle  \omega^n \rangle\!\rangle\vert_g.
\ee
Here the symbol $\langle\!\langle \omega^n \rangle\!\rangle\vert_g$ indicates the $n$-th cumulant of the pseudodistribution function $\tilde g$.
The first two are:
\begin{eqnarray}
\langle\!\langle  \omega \rangle\!\rangle\vert_g &=& \frac{\int_{-1}^1 d\omega \omega {\tilde g}(\omega)}{\int_{-1}^1 d\omega  {\tilde g}(\omega)}, \label{cum}\\
\langle\!\langle  \omega^2 \rangle\!\rangle\vert_g &=& \frac{\int_{-1}^1 d\omega \omega^2 {\tilde g}(\omega)}{\int_{-1}^1 d\omega  {\tilde g}(\omega)} -\left(\frac{\int_{-1}^1 d\omega \omega {\tilde g}(\omega)}{\int_{-1}^1 d\omega  {\tilde g}(\omega)}\right)^2.\nonumber
\end{eqnarray}
We stress $\tilde g$ is a pseudodistribution because while the above cumulant expansion is familiar in the theory of statistical distributions, ${\tilde g}$ is simply the Fourier transform of $g(t)$ and can be negative and complex, unlike a  probability distribution, which must be real and non-negative everywhere.  Nevertheless, this approach to the mathematics gives great insight into what is happening.

The connection to superoscillations is that we can characterize local rate of oscillation or growth at $t=0$ (other points can be considered by shifting the time variable, as will be shown later in the Appendices) by noticing it can be well-approximated by $g(t) \sim \exp(z t)$ where $z$ is a complex number whose real part is the local growth rate and the imaginary part is the local wavenumber.  Thus superoscillations and supergrowth can be characterized with the logarithmic derivative \cite{jordan2020superresolution}.
We can identify the local complex rate as
\be
z = i \la\!\la w \ra\!\ra.
\label{rate}
\ee
We say the function superoscillates or supergrows if the imaginary or real part of $z$ exceeds the band domain $[-1, 1]$.  This could never happen for a true probability distribution for the simple reason that the average of a variable between $[-1,1]$ must also be between $[-1,1]$. This indicates the fact $\tilde g$ can take on negative or complex values is critical to the phenomena of superoscillation/growth.  All higher cumulants determine other aspects of the function, such as the width of the superoscillation region.

\section{\label{ortho_poly}Orthogonal polynomials}
This section shows how choosing suitable basis functions can simplify the formalism of superoscillations/supergrowth for functions defined on  1-d real line.
\subsection{\label{legendre}Legendre polynomials}
To capture the interesting mathematics of super phenomena, we switch from a Fourier basis to an orthogonal polynomial basis. This change of perspective is helpful for several different reasons. We see from the section above that integrals of powers of the frequency are essential, so a polynomial expansion is natural in this respect. The Legendre polynomials constitute a natural choice in the bounded domain $[-1,1]$ as we will see shortly. Also, due to their completeness property, \emph{any} piecewise continuous function in $[-1,1]$ with a finite number of discontinuities, regardless of whether it is superoscillating/supergrowing, can be expressed as a series sum of Legendre polynomials.  Thus, we expand any function in the frequency domain as
\be
{\tilde g}(\omega) = \sum_{n=0}^\infty c_n P_n(\omega),
\label{g_legendre}
\ee
where $c_n$ are complex coefficients.  By Parseval's theorem, the total energy in the function is
\be
\int_{-\infty}^{\infty} |g(t)|^2 dt = \int_{-1}^{1} |{\tilde g}(\omega)|^2 d\omega = \sum_{n=0}^\infty  \frac{2 |c_n|^2}{2n+1},
\label{totpower}
\ee
where we have used the identity
\be
\int_{-1}^{1} P_n(x) P_m(x)dx = \frac{2}{2n+1} \delta_{n,m}.
\ee
We recall that $P_0(x) = 1$ and $P_1(x) = x$.  Consequently, we can use the orthogonality of the Legendre polynomials to characterize the moments (or cumulants) of the super-behavior.
The integral of the pseudodistribution is given by \footnote{Here, we assume that $\sum_{n=0}\left|c_nP_n(\omega)\right|$ converges for all $\omega$ and $\sum_{n=0}\int_{-1}^1d\omega c_nP_n(\omega)<\infty$. Therefore, the infinite sum and integral can be exchanged. We always deal with a finite number of terms in numerical implementations, making the issue of non-commutativity of these two operations inconsequential in practice.}
\be
\int_{-1}^1 d\omega  {\tilde g}(\omega) = 2 c_0,
\ee
while the first (unnormalized) moment is given by
\be
\int_{-1}^1 d\omega  \, \omega \, {\tilde g}(\omega) = \frac{2}{3} c_1.
\ee
 Recalling the superoscillation rate Eq.~\eqref{rate},  {\it any} superoscillating function with that rate must obey the condition $c_1 = - 3 i z c_0$, so the function has the expansion
\be
{\tilde g}(\omega) = c_0(1 - 3 i z \omega) + \sum_{n=2}^\infty c_n P_n(\omega).
\label{g_legendre_z}
\ee
This general result gives a privileged place to the Legendre polynomials to describe supergrowth/superoscillation on real line. Regardless of the values of the higher-order coefficients, such an expansion produces a local rate $z$ at the origin. Therefore, Eq.~\eqref{g_legendre_z} serves as a prescription for generating arbitrary superoscillatory/supergrowing functions. The higher order coefficients can help us constrain other properties (such as width) of the superoscillating/supergrowing region. 
\subsection{\label{real_space}Real space}
\begin{figure*}
    \centering    \includegraphics[width=\linewidth]{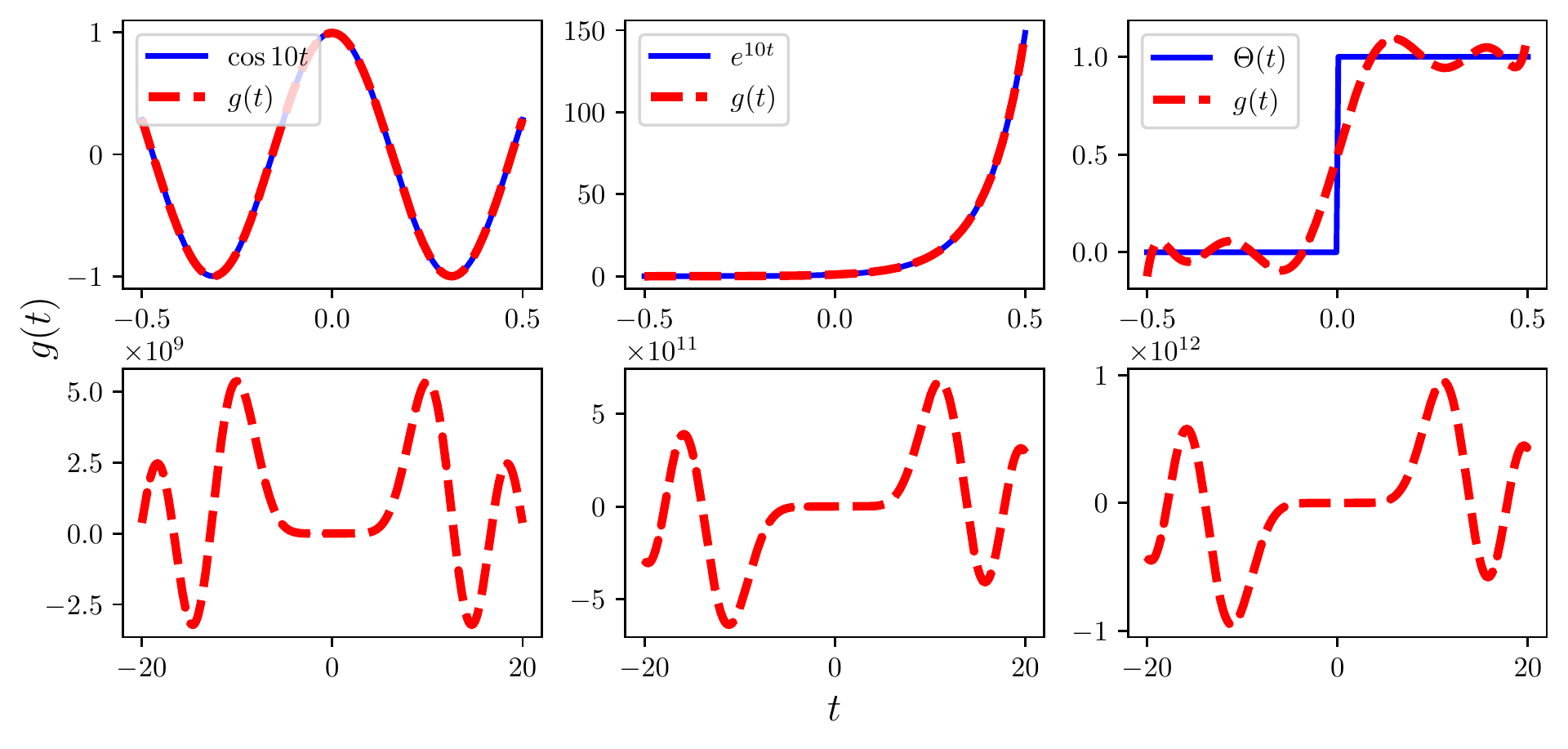}    \caption{In the top row, the left panel, the middle panel, and the right panel show in red dashed the bandlimited approximations of $\cos 10t$ (superoscillating), $e^{10t}$ (supergrowing), and unit step function $\Theta(t)$ (discontinuous), all shown as continuous blue curves,  using the method presented in Sec.~\ref{approx_real}. The interval of interest is $t\in[-1/2,1/2]$, and we only consider the first 10 terms of Eq.~\eqref{gt_bessel} for the approximation. Our method successfully reproduces the superoscillating and supergrowing functions while a continuous approximation is achieved for the step function. The bottom panels show the respective $g(t)$ in a larger interval. The smaller features between $[-1/2,1/2]$ are not visible in the bottom panels due to the extensive dynamic range involved.}
    \label{fig:approx1}
\end{figure*}
We can also Fourier transform the Legendre polynomials back into the original space to obtain the expansion\footnote{Here we use
$ j_n(t)=\tfrac{1}{2i^n}\int_{-1}^{1}d\omega e^{i\omega t}P_n(\omega).$}    
\be
g(t) = \sqrt{\frac{2}{\pi}} \sum_{n=0}^\infty i^n c_n j_n(t),
\label{gt_bessel}
\ee
where $j_n$ is the $n$-th spherical Bessel function and is related to the usual $J$-Bessel function by $j_n(t) = \sqrt{\tfrac{\pi}{2t}}J_{n+1/2}(t) $. By construction, the function $g(t)$ is  continuous.  A very nice feature of the spherical Bessel functions is they are related to the derivatives of $sinc$ function, the characteristic band-limited function via the Rayleigh formula,
\be
j_n(t) = (-t)^n \left( t^{-1} \frac{d}{dt}\right)^n {\rm sinc}(t).
\ee
This has a natural interpretation of building up the super-oscillating function as a series of individually band-limited spherical Bessel functions. Orthogonality of the spherical Bessel functions may be expressed as
\be
\int_{-\infty}^\infty j_m(t) j_n(t) dt = \frac{\pi \delta_{n,m}}{2 n+ 1},
\ee
so the total power in the signals  Eq.~\eqref{totpower} may be shown to be consistent from either space.  From our earlier expression, we can expand the function around $t=0$ that fully captures the superoscillation as
\be
g(t) = \sqrt{\frac{2}{\pi}}  \left( c_0[ j_0(t) + 3 z j_1(t)] + \sum_{n=2}^\infty i^n c_n j_n(t)\right).
\ee
The superoscillation(growth) can be increased as much as one wishes by increasing the imaginary(real) part of $z$ regardless of the higher order coefficients.

The higher-order coefficients specify the rest of the function's behavior.  Let us calculate the second (unnormalized) moment of the pseudo-distribution,
\be
\la \omega^2 \ra|_{\tilde g} =  \frac{2}{3}c_0+\frac{4}{15} c_2,
\ee
so the second cumulant in Eq.~\eqref{cum} is given by
\be
\la\!\la \omega^2 \ra\!\ra|_{\tilde g} =\frac{1}{3}+ \frac{2 c_2}{15 c_0} - \frac{c_1^2}{9 c_0^2} = \frac{1}{3}+\frac{2 c_2}{15 c_0} + z^2.
\ee
Here, we have replaced $c_1/c_0$ for the degree of supergrowth/oscillation.  In the superoscillation case, $z$ is purely imaginary, so the second cumulant can be 0 or even negative - this is possible because $\tilde g$ is a pseudodistribution.

\subsection{\label{moments}Higher order moments}
Now, we consider the $m$-th moment of $\tilde{g}(\omega)$, i.e.
\be
\Omega_m=\int_{-1}^{1}d\omega \omega^m\tilde{g}(\omega).
\label{omegamdefn}
\ee
From the definition of Fourier transform,  we see that $\Omega_m=(-i)^mg^{(m)}(0)$.  Using Eq.~\eqref{gt_bessel},  we have $g^{(m)}(0)=\sqrt{\frac{2}{\pi}}\sum_{n=0}^\infty i^n c_nj_n^{(m)}(0)$. 
Now, $j_n^{(m)}(0)=\tfrac{i^{m-n}}{2}\int_{-1}^{1}d\omega \omega^m P_n(\omega)$. Next, we use
\begin{equation}
    \omega^m=\sum_{l=m,m-2,\cdots}\frac{(2l+1)m!}{2^{(m-l)/2}(\tfrac{1}{2}(m-l))!(l+m+1)!!}P_l(\omega).
\end{equation}
This expresses $\omega^m$ as a linear combination of Legendre polynomials of orders $\le m$. Therefore, the integration $K_{m,n}=\int_{-1}^{1}d\omega\omega^mP_n(\omega) = 0$ for $n>m$ and also for $m-n=$ odd integers. For $m-n=2p$, where $p\ge0$, we have
\begin{equation}
    K_{m,n}=2\frac{m!}{2^{(m-n)/2}(\tfrac{1}{2}(m-n))!(m+n+1)!!}.
    \label{imn}
\end{equation}
Using the above results, we have 
\begin{equation}
\begin{split}
     &\Omega_m =\int_{-1}^{1}d\omega\omega^m\tilde{g}(\omega)\\&=2\sum_{n=m,m-2, \cdots}c_n \frac{m!}{2^{(m-n)/2}(\tfrac{1}{2}(m-n))!(m+n+1)!!}
\end{split}
   \label{Omega_m}
\end{equation}
Additionally, from the series expansion of spherical Bessel functions, we can write
\begin{equation}
g(t)=\sqrt{\frac{2}{\pi}}\sum_{n=0}^\infty c_n(it)^n\sum_{q=0}^\infty \frac{(-t^2/2)^q}{q!(2n+2q+1)!!},
    \label{gt_taylor}
\end{equation}
which expresses $g(t)$ in terms of powers of $t$ \cite{chremmos_superoscillations_2015,soda_efficient_2020}.

\section{\label{approx_scheme}Approximation scheme}

Here we provide a prescription to approximate an arbitrary function in an interval in terms of bandlimited functions. For completeness, we look at functions defined on the real line as well as cylindrically symmetric functions.
\subsection{\label{approx_real}Approximation on real line}

We can utilize the spherical Bessel series expansion Eq.~\eqref{gt_bessel} to approximate a function $F(t)$ between $t_i$ and $t_f$. We only consider the first $M$ terms of the series expansion. The squared integral of the approximation error in the interval is given by
\begin{equation}
\epsilon(t_i,t_f)=\int_{t_i}^{t_f}\left| f(t)-\sqrt{2/\pi}\sum_{n=0}^{M-1}\gamma_n j_n(t)\right|^2dt.
    \label{error}
\end{equation}
Extremizing this wrt $\gamma_n^\star$ leads to the matrix equation
\begin{equation}
\sum_{m=0}^{M-1}A_{nm}(t_i,t_f)\gamma_m=\sqrt{\pi/2}\int_{t_i}^{t_f}f(t)j_n(t)dt,
    \label{mateqn}
\end{equation}
for $n=0,1,2,\cdots,M-1$,  and
\be
A_{nm}(t_i,t_f)=\int_{t_i}^{t_f}j_n(t)j_m(t)dt. 
\ee
Eq.~\eqref{mateqn} can be expressed as the matrix equation $A\Gamma=B$, where $\Gamma_n=\gamma_n$ and 
\be
B_n=\sqrt{\pi/2}\int_{t_i}^{t_f}f(t)j_n(t)dt.
\ee
Given $F(t)$, we can invert $A$ and  solve for $\Gamma=A^{-1}B$ to construct an approximate function $g(t)$. Fig.~\ref{fig:approx1} shows the efficacy of this method by reproducing a superoscillating and a supergrowing function in the interval $[-1/2,1/2]$. For a discontinuous step function, a continuous  bandlimited approximation is achieved. We note that $A$ can be computed only once; different functions to approximate only change $B$.

The overlaps between spherical Bessel functions in the interval $[t_i,t_f]$  can be close to zero, leading to an ill-conditioned matrix $A$. In such cases, the Moore-Penrose pseudo-inverse of $A$, denoted as $A^+$, can be used to calculate $\Gamma=A^+B$. This method inverts non-zero singular values of $A$ above a specified threshold. Eq.~\eqref{error} can also be minimized with respect to $\gamma_n$ using  other standard numerical optimization schemes.

\begin{figure}
    \centering    \includegraphics[width=\linewidth]{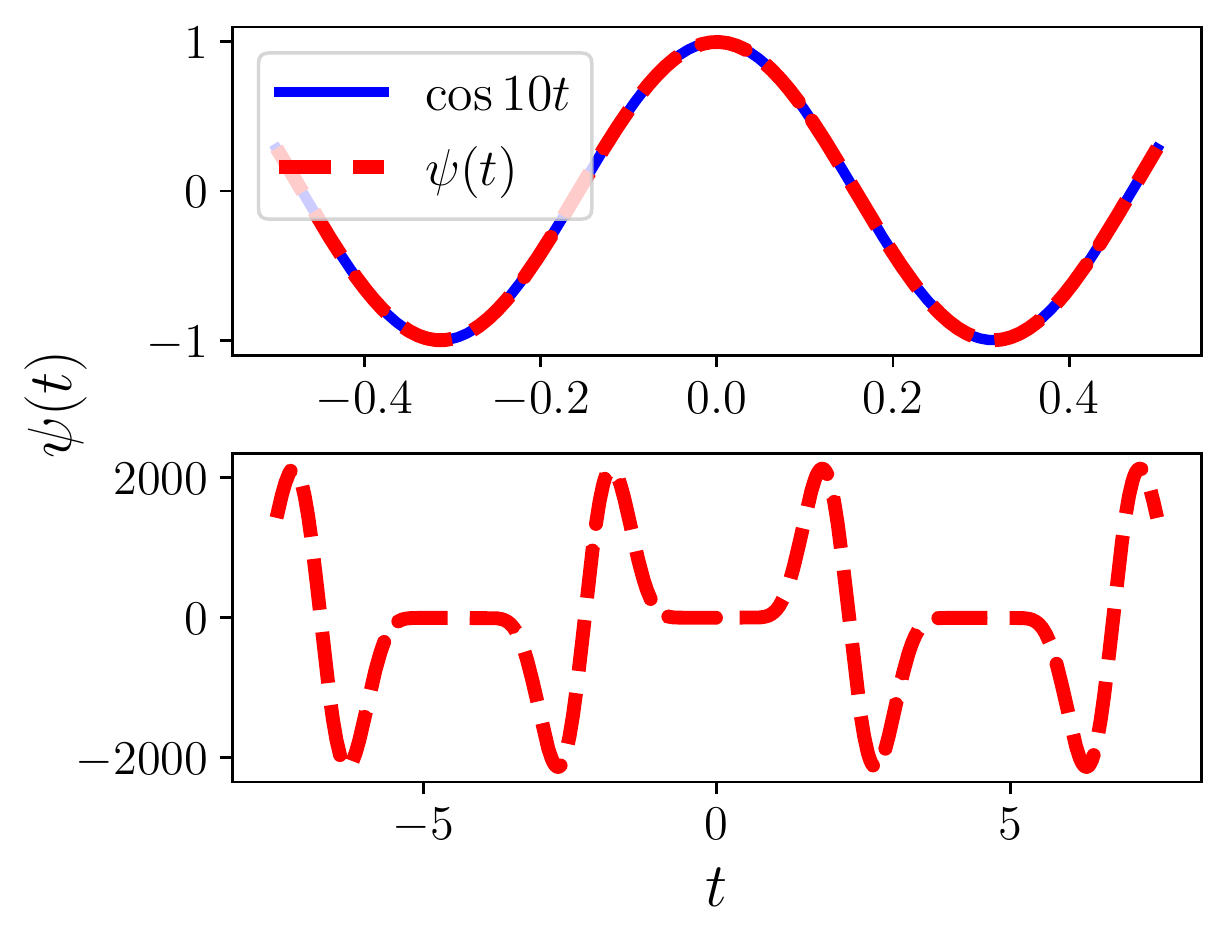}    \caption{Top panel shows the target function $\cos10t$, in blue solid and the approximating periodic function $\psi(t)=\sum_{n=0}^9C_ne^{i\omega_nt}$ in red dashed.  The interval of interest is $t\in[-1/2,1/2]$. The bottom panel shows the behaviour of the periodic approximation for a larger interval.}
    \label{fig:approxPeriodic}
\end{figure}

  Similar methods can be utilized for any periodic basis functions on a finite space interval. For example, we choose $\psi(t)=\sum_{n=0}^NC_ne^{i\omega_nt}$, with $\omega_n=(1-\tfrac{2n}{N})2\pi$. Now, the question becomes, how to choose $C_n$ such that $\psi(t)$ mimics a function $\Phi(t)$ between $t_1$ and $t_2$? Minimizing the integral 
\begin{equation}
\epsilon(t_1,t_2)=\int_{t_1}^{t_2}\left| \Phi(t)-\sum_{n=0}^{N}C_n e^{i\omega_nt}\right|^2dx
    \label{error1}
\end{equation}
wrt $C_n^\star$, we get the equations $\sum_{m=0}^{N}\alpha_{nm}C_m=b_n $, where $b_n=\int_{t_1}^{t_2}\Phi(t)e^{-i\omega_nt}dt$ and
\begin{equation}
\begin{split}
    &\alpha_{nm}=\int_{t_1}^{t_2}e^{i(\omega_m-\omega_n)t}dt=\Delta t\quad \text{for }  n=m\\ =& \Delta te^{i\frac{4\pi(n-m)}{N}\tfrac{t_1+t_2}{2}}{\rm sinc}\left(\tfrac{2\pi(n-m)\Delta t}{N}\right) \text{for } n\neq m,
\end{split}
    \label{alpha_mat}
\end{equation}
with $\Delta t=t_2-t_1$. Fig.~\ref{fig:approxPeriodic} shows  an approximation of $\cos10t$ in terms of a periodic function for $t\in[-1/2,1/2]$. 

\begin{figure*}
    \centering    \includegraphics[width=\linewidth]{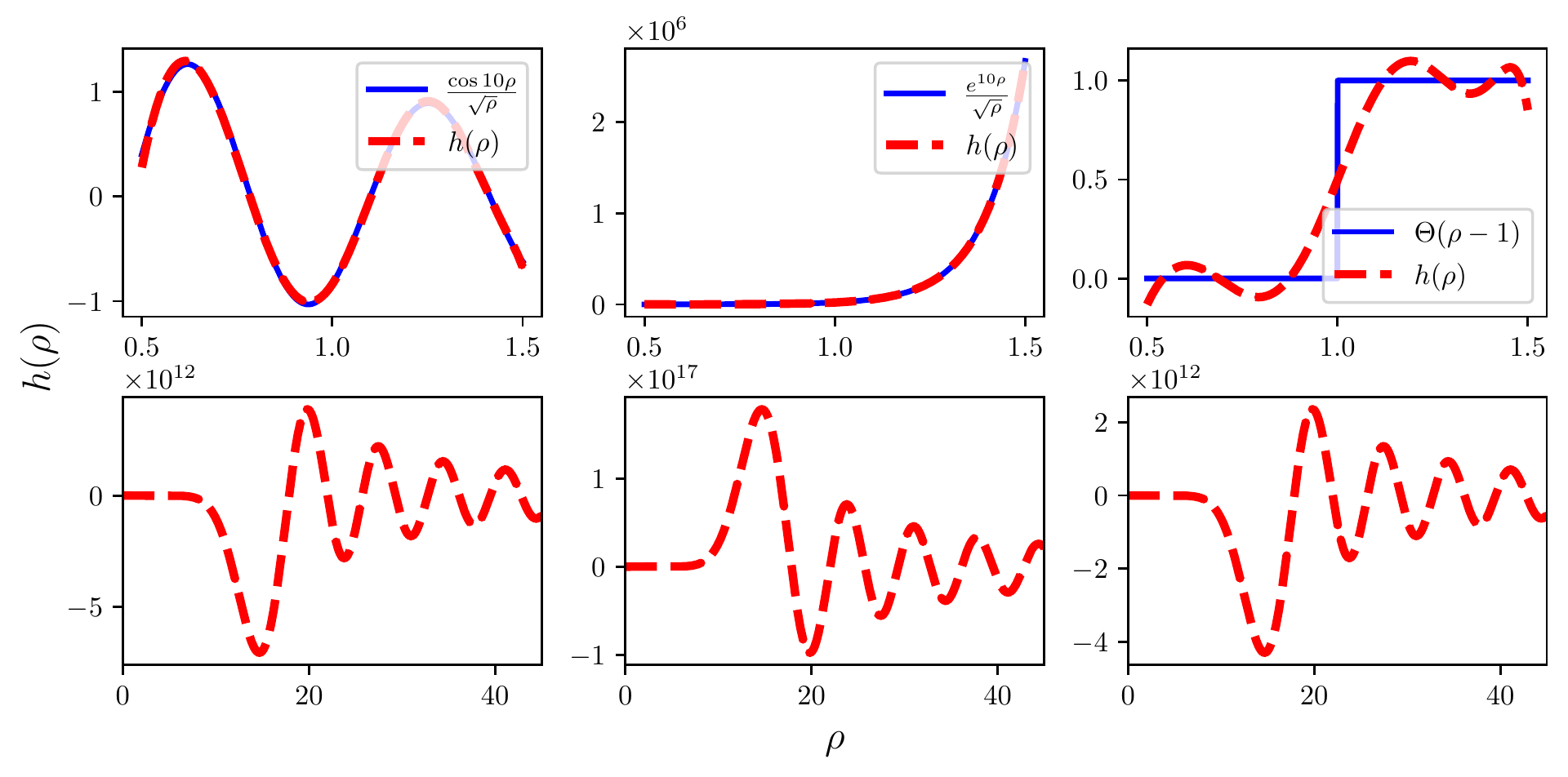}    \caption{The left panel, the middle panel, and the right panel in the top row show in red dashed the approximations of $\tfrac{\cos 10\rho}{\sqrt{\rho}}$ (superoscillating), $\tfrac{e^{10\rho}}{\sqrt{\rho}}$ (supergrowing) and unit step function $\Theta(\rho-1)$ (discontinuous), all shown as continuous blue curves,  using the method presented in Sec.~\ref{approx_cyl}. The interval of interest is the radius $\rho\in[1/2,3/2]$, and we consider the first 10 terms of Eq.~\eqref{h_bessel}. Again we see a successful reproduction of superoscillating and supergrowing functions and a continuous approximation for the radial step function. The bottom panels, once more, show the respective $h(\rho)$ in a larger interval. }
    \label{fig:approx2}
\end{figure*}

\subsection{\label{approx_cyl}Approximation for cylindrically symmetric functions}
In optical systems, 2-d cylindrical geometry  is the most relevant. Here we consider functions with cylindrical  symmetry for simplicity. Appendices \ref{cylindrical} and \ref{gen_def} describe the mathematical framework suitable for such functions. To approximate a function $H(\rho)$ in a radius $\rho\in[\rho_i,\rho_f]$ using the expansion   in Eq.~\eqref{h_bessel}, error is defined as 
\begin{equation}
\epsilon(\rho_i,\rho_f)=\int_{\rho_i}^{\rho_f}\left| H(\rho)-\sum_{n=0}^{M-1}(-1)^na_{2n} \frac{J_{2n+1}(\rho)}{\rho}\right|^2\rho d\rho,
    \label{error_rho}
\end{equation}
where $a_{2n}$  is expansion coefficient in a Bessel function $J_{2n+1}$ basis.
We can again minimize the above equation wrt the coefficients to find a suitable approximation. Fig.~\ref{fig:approx2} shows the approximations for a superoscillating, a supergrowing and a discontinuous function in this geometry. 

\section{\label{enerbound}Energy considerations}
We now investigate how the fractional energy in a specified interval of a bandlimited function can be bounded by the local rates within the interval. The energy between $t_i\le t\le t_f$ is defined as 
\begin{equation}
    \Delta E= \int_{t_i}^{t_f}dt \left|g(t)\right|^2, 
\end{equation}
 where  the total energy is 
 \begin{equation}
     E= \int_{-\infty}^{\infty}dt \left|g(t)\right|^2.
     \label{totalE}
\end{equation}
If $r$ denotes the minimum value of the logarithmic derivative of $g(t)$ in this interval, then we can show (see Appendix.~\ref{energies})
\begin{equation}
   \tfrac{\Delta E}{E}\le\frac{\Delta t}{2(1+3r^2)},
\end{equation}
 where $\Delta t=t_f-t_i$. The inequality above puts a local rate dependent upper bound on the fractional energy in a region of a given length. As expected, if we consider a highly superoscillating/supergrowing region, $r\gg 1$, the amount of energy within becomes very limited.

\section{\label{discussion}Discussion} 
We present a new perspective on superoscillatory/supergrowing functions based on their behavior in Fourier space. We show that, for such functions, the Fourier transform can be interpreted as a pseudodistribution. The cumulants of the pseudodistribution correspond to the logarithmic derivatives of the original function. We then use Legendre polynomials to describe 1-d bandlimited functions on a real line. The first two coefficients of the Legendre polynomial expansion are sufficient to constrain the superoscillatory/supergrowing at the origin. These results offer a way to generate functions with arbitrary local behavior. Returning to the original space, the function can be expressed as a series of spherical Bessel functions. This allows us to formulate a scheme to construct bandlimited functions that mimic a desirable behavior in a specified interval. Our formalism successfully applies to cylindrically symmetric functions as well and can be extended to any geometry. Finally, we attain a local rate dependent bound on the fractional energy content in an interval.

Our results offer a multitude of potential applications.  The Fourier space based description could help us construct supergrowing/superoscillating functions in the frequency space. Also, both the Legendre and spherical Bessel series expansions grant us precise and tractable control over the superoscillatory/supergrowing properties of the function.  Besides, the approximation scheme serves as the groundwork for numerical supergrowing/superoscillating optical field spot generation and optimization techniques. Analyzing cylindrically symmetric functions makes the results immediately relevant for diffraction-limited imaging systems in laboratory settings. Additionally, the bound on fractional energy estimates the necessary dynamic range in experimental scenarios. Furthermore, the appendices generalize the Legendre/spherical Bessel series expansion to arbitrary points, i.e.~away from the origin. We also include analytical results on the energy content in an interval and correlation relations of the spherical Bessel functions. These investigations can be helpful for an in-depth look at the properties of bandlimited functions on real lines.

The insights we provide can be extended in several aspects. Future possible investigations include generalizing the results to different (non-cylindrically symmetric, spherical) geometries and higher dimensions. Also, the approximation scheme presented here deals with a single interval. Extending it for multiple disjoint intervals can help us design functions with conveniently placed supergrowing and superoscillatory regions. Jordan et al.~\cite{jordanSuperphenomenaArbitraryQuantum2022} generalized the super phenomena to arbitrary quantum observables. It will be interesting to see whether a formalism like the one presented here can be applied to analyze the weak values of quantum observables. Although the benefits of supergrowth based superresolution imaging are known,  supergrowing functions are yet to be realized experimentally. An imminent challenge is to design supergrowing functions with a dynamic range that can be implemented in state of the art imaging setups.  Our results provide a foundation for such ventures and therefore, serve as a methodical framework for superoscillation/supergrowth based superresolution  imaging.
 
 \section*{Acknowledgements}
 We thank Yakir Aharonov, Sandu Popescu, Anurag Sahay,  Nick Vamivakas, Sethuraj Karimparambil Raju, Abhishek Chakraborty,  Sultan Abdul Wadood, John Howell, Achim Kempf, Daniele C.~Struppa, Fabrizio Colombo, Irene Sabadini for providing insight throughout the project. 
This work was supported by the AFOSR grant \#FA9550-21-1-0322 and the Bill Hannon Foundation.
\appendix
\section{\label{so_everywhere}Generating a superoscillation everywhere}
 We can apply the method of the Taylor series approach to generate a desired behavior.  For example, if we want to produce a superoscillation everywhere \cite{aharonov2021new}, $F(t) = e^{i s t}$, where $s>1$, we can determine the expansion coefficients successively via the Taylor series up to first $M$ terms,
\be
(i s)^N = \sqrt{\frac{2}{\pi}}\sum_{n = 0}^{M-1} i^n c_n j_n^{(N)}(0).
\ee
Here $j_n^{(N)}(0)$ is the $N$th derivative of the $n$th spherical Bessel function, evaluated at $t=0$.  The coefficients can then be determined successively.  For example, if $N=0$, we find $c_0=\sqrt{\pi/2}$, for $N=1$, we find $c_1 = 3s\sqrt{\pi/2}$. for $N=2$, we find $c_2 = \tfrac{15}{2}(s^2-1/3)\sqrt{\pi/2}$, and so on.  Generally, $c_n$ scales as $s^n$ for large $n$ plus corrections of a smaller order.  For $s>1$, convergence is doubtful, but we can always truncate at finite $N$ to mimic the behavior for a large but finite $t$. The method works in finite domain case as well.

\section{\label{general_origin}Superoscillatory properties at a general coordinate}
To analyze the superoscillatory properties of the function $g(t)$ in Eq.~\eqref{gt_bessel} at a general coordinate $t=t'$, we can shift its origin to $t'$. This will in general lead to a different set of coefficients.  We can express this fact by considering an expansion of the type
\begin{equation}
    g(t)=\sqrt{\frac{2}{\pi}}\sum_{n=0}^\infty \gamma_n(t^\prime)j_n(t-t^\prime), 
\label{general_expansion}
\end{equation}
where  $\gamma_n$ are related to the coefficients in Eq.~\eqref{gt_bessel} as $\gamma_n(0)=i^nc_n$. The coefficients are functions of $t'$ because their values depend on the choice of the origin of expansion. 

In this case $\frac{g^\prime(t^\prime)}{g(t^\prime)}=\frac{\gamma_1(t^\prime)}{3\gamma_0(t^\prime)}$. That is, by expressing the function in terms of shifted spherical Bessel functions, we can find out the logarithmic derivative at any desirable location. Now, using the orthonormality of spherical Bessel functions, it can be shown that
\begin{equation}
    \gamma_n(t^\prime)=\tfrac{2n+1}{\sqrt{2\pi}}\int_{-\infty}^\infty ds g(s+t^\prime)j_n(s).
    \label{gammas}
\end{equation}

Since the convolution of a (bandlimited) function with $j_0$ returns the same function, $\gamma_0(t^\prime)=\sqrt{\frac{\pi}{2}}g(t^\prime)$. 

For any shifted expansion, the function $g(t)$ should only depend on $t$, not $t^\prime$. Therefore the coefficients must satisfy 
\begin{equation}
\begin{split}
     &\partial_{t^\prime}g(t)=\sqrt{\frac{2}{\pi}}\sum_{n=0}^\infty j_n(t-t^\prime)\times\\&\left(\gamma_n^\prime(t^\prime)-\tfrac{n+1}{2n+3}\gamma_{n+1}(t^\prime)+\tfrac{n}{2n-1}\gamma_{n-1}(t^\prime)\right) =0 ,
\end{split}
   \label{tp_der}
\end{equation}
where, we have used $n j_{n-1}-(n+1)j_{n+1}=(2n+1)j_n^\prime$. Thus, for a general expansion like Eq.~\eqref{general_expansion} to produce a single variable bandlimited function, the coefficients should satisfy the recurrence relations
\begin{equation}
    \gamma_{n+1}(t^\prime)=\frac{2n+3}{n+1}\left(\gamma_n^\prime(t^\prime)+\tfrac{n}{2n-1}\gamma_{n-1}(t^\prime)\right).
    \label{recurrence}
\end{equation}
This recurrence relation can be used to construct a desired function $g(t)$.

For example, we can use the recurrence identity of the spherical Bessel functions shown above, and define $\gamma_n(t^\prime)=(2n+1)(-1)^nj_n(t^\prime)$. It is easy to see that such coefficients satisfy Eq.~\eqref{recurrence}. Using such coefficients in Eq.~\eqref{general_expansion} and using the identity $\sum_{n=0}^\infty(2n+1)j_n(t_1)j_n(t_2)=\frac{\sin(t_1-t_2)}{t_1-t_2}$ we get $g(t)\sim \frac{\sin t}{t}\sim j_0(t)$. This also matches with our observation of $\gamma_0(t)=g(t)$.

An alternative approach is to look at the Fourier space at $t=t^\prime$

\be
g(t^\prime)=\int_{-1}^{1}\frac{d\omega}{\sqrt{2\pi}}e^{i\omega t^\prime}\tilde{g}(\omega).
\ee
The $n^{\textrm{th}}$ derivative of $g(t^\prime)$ is

\be
g^{(n)}(t^\prime)=\int_{-1}^{1}\frac{d\omega}{\sqrt{2\pi}}(i\omega)^ne^{i\omega t^\prime}\tilde{g}(\omega).
\ee
Additionally, we can use the Maclaurin series expansion of $e^{i\omega t^\prime}$

\be
e^{i\omega t^\prime}=\sum_{m=0}^\infty (i\omega)^m\frac{t^{\prime m}}{m!}.
\ee
Therefore,
\be
\begin{split}
   g^{(n)}(t^\prime)&=\int_{-1}^{1}\frac{d\omega}{\sqrt{2\pi}}(i\omega)^n\sum_{m=0}^\infty (i\omega)^m\frac{t^{\prime m}}{m!}\tilde{g}(\omega),\\&=  \sum_{m=0}^\infty \frac{t^{\prime m}}{m!}\frac{i^{n+m}\Omega_{n+m}}{\sqrt{2\pi}},
   \label{srsexpn}
\end{split}
\ee
where we use the definition in Eq.~\eqref{omegamdefn}.
We have assumed the sum and the integral to be interchangeable. Eq.~\eqref{srsexpn} expresses $g(t)$ and all of its higher derivatives as a Maclaurin series at an arbitrary coordinate $t=t^\prime$.
\section{\label{correlation}Correlation functions}
We define the correlation function
\begin{equation}
    I_{m,n}(t)=\int_{-\infty}^{\infty}dsj_m(s+t)j_n(s).
    \label{Imn}
\end{equation}
It is easy to see that $I_{m,n}(t)=I_{n,m}(-t)$, and 
\begin{equation}
    I_{m+1,n}(t)=\tfrac{m}{m+1}I_{m-1,n}(t)-\tfrac{2m+1}{m+1}I_{m,n}^\prime(t).
    \label{recurrenceI}
\end{equation} 
Also, $I_{0,n}(t)=\pi (-1)^nj_n(t)$, and $I_{1,n}(t)=\pi (-1)^{n+1}j_n^\prime(t)$. These two identities, along with Eq.~\eqref{recurrenceI}, can be used to find any $I_{m,n}(t)$. Using these second-order correlation functions, we can express shifted spherical Bessel function as 
\begin{equation}
j_m(s+t)=\sum_{n=0}^\infty \tfrac{2n+1}{\pi}I_{m,n}(t)j_n(s).
    \label{shiftedjm}
\end{equation}
 Using Eqs.~\eqref{general_expansion},  \eqref{gammas} and \eqref{shiftedjm}, we find
 \begin{equation}
     \gamma_m(t^\prime)=\tfrac{2m+1}{\pi}\sum_{p=0}^\infty \gamma_p(0)I_{p,m}(t^\prime).
     \label{gammatgamma0}
 \end{equation}
On the other hand,
\begin{equation}
\gamma_p(0)=\tfrac{2p+1}{\pi}\sum_{m=0}^\infty \gamma_m(t^\prime)I_{p,m}(t^\prime).
     \label{gamma0gammat}
 \end{equation}
 These identities connect the coefficients (local growth rate/wave number in turn) at an arbitrary point to the coefficients (local rates) at the origin.

\section{\label{energies}Energy calculation}
 We can integrate the absolute square of Eq.~\eqref{general_expansion} to get the total energy content of the function $g(t)$
 \begin{equation}
     E=\int_{-\infty}^{\infty}dt|g(t)|^2=\sum_{n=0}^{\infty}\frac{2|\gamma_n(t^\prime)|^2}{2n+1}.
     \label{tot_E}
 \end{equation}
It should be noted that this expression is independent of $t^\prime$. We can also define energy until $t=T$ 
\begin{equation}
    E(T)=\int_{-\infty}^{T}dt|g(t)|^2.
    \label{ET}
\end{equation}
Using Eq.~\eqref{general_expansion} for $t'=T$, it  can be shown that $ E(T)=$
\begin{equation}
   \tfrac{E}{2}-\tfrac{4}{\pi}\sum_{n,p=0}^\infty \text{Re}\left(\gamma_n(T)\gamma_{n+2p+1}^\star(T)\right)\int_0^\infty dt j_n(t)j_{n+2p+1}(t).
    \label{ET_expanse}
\end{equation}
We use the result from Bessel integrals $\int_0^\infty  dt j_n(t)j_{n+2p+1}(t)=\frac{(-1)^p}{2(2p+1)(n+p+1)}$. Thus, $E(T)=$
\begin{equation}
    \tfrac{E}{2}-\tfrac{4}{\pi}\sum_{n,p=0}^\infty \text{Re}\left(\gamma_n(T)\gamma_{n+2p+1}^\star(T)\right)\frac{(-1)^p}{2(2p+1)(n+p+1)}.
    \label{ET_expanse1}
\end{equation}
Using the expression above we can calculate the intensity between points $t_i$ and $t_f$ as $E(t_i,t_f)=E(t_f)-E(t_i)$. Further,  we can use Eq.~\eqref{gammatgamma0} to express $E(T)$ in terms of coefficients at the origin if necessary.

Now define the energy in the interval $(t_i,t_f)$ to be $\Delta E=E(t_i,t_f)$. Then 
\begin{equation}
    \tfrac{\Delta E}{E}=\frac{\int_{t_i}^{t_f}dt|g(t)|^2}{\sum_{n=0}^{\infty}\tfrac{2|\gamma_n(t^\prime)|^2}{2n+1}},
\end{equation}
where we assume $t_i\le t'\le t_f$. Assuming the logarithmic derivative $\left|\tfrac{\gamma_1(t')}{3\gamma_0(t')}\right|\ge r$ in this interval,
 we have $E\ge2|\gamma_0(t')|^2(1+3r^2)$. Therefore, 
 \begin{equation}
    \tfrac{\Delta E}{E}\le\frac{1}{2(1+3r^2)}\int_{t_i}^{t_f}dt \left|\tfrac{g(t)}{\gamma_0(t')}\right|^2.
    \label{dE_E1}
\end{equation}
 We can choose $t'$ such that $|\gamma_0(t')|=\max_{t_i\le t\le t_f} |g(t)|. $ This gives us, 
 \begin{equation}
   \tfrac{\Delta E}{E}\le\frac{\Delta t}{2(1+3r^2)},
\end{equation}
 where $\Delta t=t_f-t_i$. 

\section{\label{cylindrical}Cylindrical geometry}
 We consider the case of cylindrically symmetric 2d functions. In this case, the Fourier transform also is cylindrically symmetric and is the same as the Hankel transform,
 \begin{equation}
     \tilde{h}(\nu)=\int_0^{\infty}d\rho \rho h(\rho) J_0(\nu\rho),
     \label{hankel_transform}
 \end{equation}
 and the inverse transform is 
 \begin{equation}
     h(\rho)=\int_0^{\infty}d\nu \nu \tilde{h}(\nu) J_0(\nu\rho).
     \label{ihankel_transform}
 \end{equation}
 We only consider functions bandlimited on the unit disk, i.e., $\tilde{h}(\nu)=0$ for $\nu>1$. 

 Similar to the $1$-d case, the function $\tilde{h}(\nu)$ could be expressed in terms of Zernike polynomials \cite{Cerjan:07} on the unit disk.
 \begin{equation}
     \tilde{h}(\nu)=\sum_{n=0}^\infty a_{2n} R_{2n}^0(\nu),
     \label{ht_zernike}
 \end{equation}
 with $a_{2n}$ expansion coefficients 
and the Zernike polynomials are
 \begin{equation}
     R_n^l(\nu)= \begin{cases}
             \sum_{p=0}^{\tfrac{n-l}{2}} \frac{(-1)^p(n-p)!}{p!(\tfrac{n+l}{2}-p)!(\tfrac{n-l}{2}-p)!}\nu^{n-2p}  &  n-l=\text{even} \\
             0  &  n-l=\text{odd}
       \end{cases} 
     \label{zernike}
 \end{equation}
 They satisfy the orthogonality 
 \begin{equation}
     \int_0^1d\nu\nu  R_n^l(\nu)  R_{n^\prime}^l(\nu)=\frac{\delta_{n{n^\prime}}}{2n+1}. 
 \end{equation}
 The polynomials Hankel transform into Bessel functions. Thus, in real space, we have 
 \begin{equation}
     h(\rho)=\sum_{n=0}^\infty (-1)^na_{2n} \frac{J_{2n+1}(\rho)}{\rho}.
     \label{h_bessel}
 \end{equation}

 \section{\label{gen_def}Generalized definitions of local rates}
 Since cylindrically symmetric waves away from the origin take the form $\frac{e^{ik\rho}}{\sqrt{\rho}}$, we see that a definition of local rates in terms of the logarithmic derivative of $h(\rho)$ might be incomplete. Instead, by inverting the Helmholtz equation, we propose a new  definition of local wave number k and local growth rate $\kappa$ as \cite{jordan2020superresolution}
 \begin{equation}
(\kappa+ik)^2=\frac{\nabla^2h}{h}.
     \label{newdef}
 \end{equation}
 If $h_r$ and $h_i$ denote the real and imaginary parts of $h$ respectively, then,
 \begin{equation}
     \begin{split}
         \kappa^2 &= \tfrac{1}{2}\left[A+\sqrt{A^2+B^2}\right],\\
         k^2 &= \tfrac{1}{2}\left[-A+\sqrt{A^2+B^2}\right],
     \end{split}
 \end{equation}
 where 
 \begin{equation}
     \begin{split}
         A &= \frac{h_r\nabla^2h_r+h_i\nabla^2h_i}{|h|^2},\\
         B &= \frac{h_r\nabla^2h_i-h_i\nabla^2h_r}{|h|^2}.
     \end{split}
 \end{equation}
\bibliography{refs}

\begin{thebibliography}{22}%
\makeatletter
\providecommand \@ifxundefined [1]{%
 \@ifx{#1\undefined}
}%
\providecommand \@ifnum [1]{%
 \ifnum #1\expandafter \@firstoftwo
 \else \expandafter \@secondoftwo
 \fi
}%
\providecommand \@ifx [1]{%
 \ifx #1\expandafter \@firstoftwo
 \else \expandafter \@secondoftwo
 \fi
}%
\providecommand \natexlab [1]{#1}%
\providecommand \enquote  [1]{``#1''}%
\providecommand \bibnamefont  [1]{#1}%
\providecommand \bibfnamefont [1]{#1}%
\providecommand \citenamefont [1]{#1}%
\providecommand \href@noop [0]{\@secondoftwo}%
\providecommand \href [0]{\begingroup \@sanitize@url \@href}%
\providecommand \@href[1]{\@@startlink{#1}\@@href}%
\providecommand \@@href[1]{\endgroup#1\@@endlink}%
\providecommand \@sanitize@url [0]{\catcode `\\12\catcode `\$12\catcode
  `\&12\catcode `\#12\catcode `\^12\catcode `\_12\catcode `\%12\relax}%
\providecommand \@@startlink[1]{}%
\providecommand \@@endlink[0]{}%
\providecommand \url  [0]{\begingroup\@sanitize@url \@url }%
\providecommand \@url [1]{\endgroup\@href {#1}{\urlprefix }}%
\providecommand \urlprefix  [0]{URL }%
\providecommand \Eprint [0]{\href }%
\providecommand \doibase [0]{https://doi.org/}%
\providecommand \selectlanguage [0]{\@gobble}%
\providecommand \bibinfo  [0]{\@secondoftwo}%
\providecommand \bibfield  [0]{\@secondoftwo}%
\providecommand \translation [1]{[#1]}%
\providecommand \BibitemOpen [0]{}%
\providecommand \bibitemStop [0]{}%
\providecommand \bibitemNoStop [0]{.\EOS\space}%
\providecommand \EOS [0]{\spacefactor3000\relax}%
\providecommand \BibitemShut  [1]{\csname bibitem#1\endcsname}%
\let\auto@bib@innerbib\@empty
\bibitem [{\citenamefont {Berry}\ \emph {et~al.}(2019)\citenamefont {Berry},
  \citenamefont {Zheludev}, \citenamefont {Aharonov}, \citenamefont {Colombo},
  \citenamefont {Sabadini}, \citenamefont {Struppa}, \citenamefont {Tollaksen},
  \citenamefont {Rogers}, \citenamefont {Qin}, \citenamefont {Hong} \emph
  {et~al.}}]{berry2019roadmap}%
  \BibitemOpen
  \bibfield  {author} {\bibinfo {author} {\bibfnamefont {M.}~\bibnamefont
  {Berry}}, \bibinfo {author} {\bibfnamefont {N.}~\bibnamefont {Zheludev}},
  \bibinfo {author} {\bibfnamefont {Y.}~\bibnamefont {Aharonov}}, \bibinfo
  {author} {\bibfnamefont {F.}~\bibnamefont {Colombo}}, \bibinfo {author}
  {\bibfnamefont {I.}~\bibnamefont {Sabadini}}, \bibinfo {author}
  {\bibfnamefont {D.~C.}\ \bibnamefont {Struppa}}, \bibinfo {author}
  {\bibfnamefont {J.}~\bibnamefont {Tollaksen}}, \bibinfo {author}
  {\bibfnamefont {E.~T.}\ \bibnamefont {Rogers}}, \bibinfo {author}
  {\bibfnamefont {F.}~\bibnamefont {Qin}}, \bibinfo {author} {\bibfnamefont
  {M.}~\bibnamefont {Hong}}, \emph {et~al.},\ }\bibfield  {title} {\bibinfo
  {title} {Roadmap on superoscillations},\ }\href
  {https://doi.org/10.1088/2040-8986/ab0191} {\bibfield  {journal} {\bibinfo
  {journal} {Journal of Optics}\ }\textbf {\bibinfo {volume} {21}},\ \bibinfo
  {pages} {053002} (\bibinfo {year} {2019})}\BibitemShut {NoStop}%
\bibitem [{\citenamefont {Jordan}(2020)}]{jordan2020superresolution}%
  \BibitemOpen
  \bibfield  {author} {\bibinfo {author} {\bibfnamefont {A.~N.}\ \bibnamefont
  {Jordan}},\ }\bibfield  {title} {\bibinfo {title} {Superresolution using
  supergrowth and intensity contrast imaging},\ }\href
  {https://doi.org/10.1007/s40509-019-00214-5} {\bibfield  {journal} {\bibinfo
  {journal} {Quantum Studies: Mathematics and Foundations}\ }\textbf {\bibinfo
  {volume} {7}},\ \bibinfo {pages} {285} (\bibinfo {year} {2020})}\BibitemShut
  {NoStop}%
\bibitem [{\citenamefont {Karmakar}\ \emph {et~al.}(2023)\citenamefont
  {Karmakar}, \citenamefont {Chakraborty}, \citenamefont {Vamivakas},\ and\
  \citenamefont {Jordan}}]{SG_imaging_drafft}%
  \BibitemOpen
  \bibfield  {author} {\bibinfo {author} {\bibfnamefont {T.}~\bibnamefont
  {Karmakar}}, \bibinfo {author} {\bibfnamefont {A.}~\bibnamefont
  {Chakraborty}}, \bibinfo {author} {\bibfnamefont {A.~N.}\ \bibnamefont
  {Vamivakas}},\ and\ \bibinfo {author} {\bibfnamefont {A.~N.}\ \bibnamefont
  {Jordan}},\ }\bibfield  {title} {\bibinfo {title} {Supergrowth and
  sub-wavelength object imaging},\ }\href@noop {} {\bibfield  {journal}
  {\bibinfo  {journal} {In preparation.}\ } (\bibinfo {year}
  {2023})}\BibitemShut {NoStop}%
\bibitem [{\citenamefont {Chen}\ \emph {et~al.}(2019)\citenamefont {Chen},
  \citenamefont {Wen},\ and\ \citenamefont {Qiu}}]{chen_superoscillation_2019}%
  \BibitemOpen
  \bibfield  {author} {\bibinfo {author} {\bibfnamefont {G.}~\bibnamefont
  {Chen}}, \bibinfo {author} {\bibfnamefont {Z.-Q.}\ \bibnamefont {Wen}},\ and\
  \bibinfo {author} {\bibfnamefont {C.-W.}\ \bibnamefont {Qiu}},\ }\bibfield
  {title} {\bibinfo {title} {Superoscillation: from physics to optical
  applications},\ }\href {https://doi.org/10.1038/s41377-019-0163-9} {\bibfield
   {journal} {\bibinfo  {journal} {Light: Science \& Applications}\ }\textbf
  {\bibinfo {volume} {8}},\ \bibinfo {pages} {56} (\bibinfo {year}
  {2019})}\BibitemShut {NoStop}%
\bibitem [{\citenamefont {Baumgartl}\ \emph {et~al.}(2011)\citenamefont
  {Baumgartl}, \citenamefont {Kosmeier}, \citenamefont {Mazilu}, \citenamefont
  {Rogers}, \citenamefont {Zheludev},\ and\ \citenamefont
  {Dholakia}}]{baumgartl_far_2011}%
  \BibitemOpen
  \bibfield  {author} {\bibinfo {author} {\bibfnamefont {J.}~\bibnamefont
  {Baumgartl}}, \bibinfo {author} {\bibfnamefont {S.}~\bibnamefont {Kosmeier}},
  \bibinfo {author} {\bibfnamefont {M.}~\bibnamefont {Mazilu}}, \bibinfo
  {author} {\bibfnamefont {E.~T.~F.}\ \bibnamefont {Rogers}}, \bibinfo {author}
  {\bibfnamefont {N.~I.}\ \bibnamefont {Zheludev}},\ and\ \bibinfo {author}
  {\bibfnamefont {K.}~\bibnamefont {Dholakia}},\ }\bibfield  {title} {\bibinfo
  {title} {Far field subwavelength focusing using optical eigenmodes},\ }\href
  {https://doi.org/10.1063/1.3587636} {\bibfield  {journal} {\bibinfo
  {journal} {Applied Physics Letters}\ }\textbf {\bibinfo {volume} {98}},\
  \bibinfo {pages} {181109} (\bibinfo {year} {2011})}\BibitemShut {NoStop}%
\bibitem [{\citenamefont {Kozawa}\ and\ \citenamefont
  {Sato}(2015)}]{kozawa_numerical_2015}%
  \BibitemOpen
  \bibfield  {author} {\bibinfo {author} {\bibfnamefont {Y.}~\bibnamefont
  {Kozawa}}\ and\ \bibinfo {author} {\bibfnamefont {S.}~\bibnamefont {Sato}},\
  }\bibfield  {title} {\bibinfo {title} {Numerical analysis of resolution
  enhancement in laser scanning microscopy using a radially polarized beam},\
  }\href {https://doi.org/10.1364/OE.23.002076} {\bibfield  {journal} {\bibinfo
   {journal} {Optics Express}\ }\textbf {\bibinfo {volume} {23}},\ \bibinfo
  {pages} {2076} (\bibinfo {year} {2015})}\BibitemShut {NoStop}%
\bibitem [{\citenamefont {Kozawa}\ \emph {et~al.}(2018)\citenamefont {Kozawa},
  \citenamefont {Matsunaga},\ and\ \citenamefont {Sato}}]{Kozawa2018Feb}%
  \BibitemOpen
  \bibfield  {author} {\bibinfo {author} {\bibfnamefont {Y.}~\bibnamefont
  {Kozawa}}, \bibinfo {author} {\bibfnamefont {D.}~\bibnamefont {Matsunaga}},\
  and\ \bibinfo {author} {\bibfnamefont {S.}~\bibnamefont {Sato}},\ }\bibfield
  {title} {\bibinfo {title} {{Superresolution imaging via superoscillation
  focusing of a radially polarized beam}},\ }\href
  {https://doi.org/10.1364/OPTICA.5.000086} {\bibfield  {journal} {\bibinfo
  {journal} {Optica}\ }\textbf {\bibinfo {volume} {5}},\ \bibinfo {pages} {86}
  (\bibinfo {year} {2018})}\BibitemShut {NoStop}%
\bibitem [{\citenamefont {Diao}\ \emph {et~al.}(2016)\citenamefont {Diao},
  \citenamefont {Yuan}, \citenamefont {Yu}, \citenamefont {Zhu},\ and\
  \citenamefont {Wu}}]{diao_controllable_2016}%
  \BibitemOpen
  \bibfield  {author} {\bibinfo {author} {\bibfnamefont {J.}~\bibnamefont
  {Diao}}, \bibinfo {author} {\bibfnamefont {W.}~\bibnamefont {Yuan}}, \bibinfo
  {author} {\bibfnamefont {Y.}~\bibnamefont {Yu}}, \bibinfo {author}
  {\bibfnamefont {Y.}~\bibnamefont {Zhu}},\ and\ \bibinfo {author}
  {\bibfnamefont {Y.}~\bibnamefont {Wu}},\ }\bibfield  {title} {\bibinfo
  {title} {Controllable design of super-oscillatory planar lenses for
  sub-diffraction-limit optical needles},\ }\href
  {https://doi.org/10.1364/OE.24.001924} {\bibfield  {journal} {\bibinfo
  {journal} {Optics Express}\ }\textbf {\bibinfo {volume} {24}},\ \bibinfo
  {pages} {1924} (\bibinfo {year} {2016})}\BibitemShut {NoStop}%
\bibitem [{\citenamefont {Rogers}\ \emph {et~al.}(2018)\citenamefont {Rogers},
  \citenamefont {Bourdakos}, \citenamefont {Yuan}, \citenamefont {Mahajan},\
  and\ \citenamefont {Rogers}}]{rogersOptimisingSuperoscillatorySpots2018}%
  \BibitemOpen
  \bibfield  {author} {\bibinfo {author} {\bibfnamefont {K.~S.}\ \bibnamefont
  {Rogers}}, \bibinfo {author} {\bibfnamefont {K.~N.}\ \bibnamefont
  {Bourdakos}}, \bibinfo {author} {\bibfnamefont {G.~H.}\ \bibnamefont {Yuan}},
  \bibinfo {author} {\bibfnamefont {S.}~\bibnamefont {Mahajan}},\ and\ \bibinfo
  {author} {\bibfnamefont {E.~T.~F.}\ \bibnamefont {Rogers}},\ }\bibfield
  {title} {\bibinfo {title} {Optimising superoscillatory spots for far-field
  super-resolution imaging},\ }\href {https://doi.org/10.1364/OE.26.008095}
  {\bibfield  {journal} {\bibinfo  {journal} {Optics Express}\ }\textbf
  {\bibinfo {volume} {26}},\ \bibinfo {pages} {8095} (\bibinfo {year}
  {2018})}\BibitemShut {NoStop}%
\bibitem [{\citenamefont {Hu}\ \emph {et~al.}(2021)\citenamefont {Hu},
  \citenamefont {Wang}, \citenamefont {Jia}, \citenamefont {Fu}, \citenamefont
  {Yin}, \citenamefont {Li},\ and\ \citenamefont {Chen}}]{hu_optical_2021}%
  \BibitemOpen
  \bibfield  {author} {\bibinfo {author} {\bibfnamefont {Y.}~\bibnamefont
  {Hu}}, \bibinfo {author} {\bibfnamefont {S.}~\bibnamefont {Wang}}, \bibinfo
  {author} {\bibfnamefont {J.}~\bibnamefont {Jia}}, \bibinfo {author}
  {\bibfnamefont {S.}~\bibnamefont {Fu}}, \bibinfo {author} {\bibfnamefont
  {H.}~\bibnamefont {Yin}}, \bibinfo {author} {\bibfnamefont {Z.}~\bibnamefont
  {Li}},\ and\ \bibinfo {author} {\bibfnamefont {Z.}~\bibnamefont {Chen}},\
  }\bibfield  {title} {\bibinfo {title} {Optical superoscillatory waves without
  side lobes along a symmetric cut},\ }\href
  {https://doi.org/10.1117/1.AP.3.4.045002} {\bibfield  {journal} {\bibinfo
  {journal} {Advanced Photonics}\ }\textbf {\bibinfo {volume} {3}},\ \bibinfo
  {pages} {045002} (\bibinfo {year} {2021})}\BibitemShut {NoStop}%
\bibitem [{\citenamefont {Katzav}\ and\ \citenamefont
  {Schwartz}(2013)}]{36243494}%
  \BibitemOpen
  \bibfield  {author} {\bibinfo {author} {\bibfnamefont {E.}~\bibnamefont
  {Katzav}}\ and\ \bibinfo {author} {\bibfnamefont {M.}~\bibnamefont
  {Schwartz}},\ }\bibfield  {title} {\bibinfo {title} {Yield-optimized
  superoscillations},\ }\href {https://doi.org/10.1109/TSP.2013.2258018}
  {\bibfield  {journal} {\bibinfo  {journal} {IEEE Transactions on Signal
  Processing}\ }\textbf {\bibinfo {volume} {61}},\ \bibinfo {pages} {3113}
  (\bibinfo {year} {2013})}\BibitemShut {NoStop}%
\bibitem [{\citenamefont {Kempf}(2000)}]{kempfBlackHolesBandwidths2000}%
  \BibitemOpen
  \bibfield  {author} {\bibinfo {author} {\bibfnamefont {A.}~\bibnamefont
  {Kempf}},\ }\bibfield  {title} {\bibinfo {title} {Black holes, bandwidths and
  {{Beethoven}}},\ }\href {https://doi.org/10.1063/1.533244} {\bibfield
  {journal} {\bibinfo  {journal} {Journal of Mathematical Physics}\ }\textbf
  {\bibinfo {volume} {41}},\ \bibinfo {pages} {2360} (\bibinfo {year}
  {2000})}\BibitemShut {NoStop}%
\bibitem [{\citenamefont {Kempf}(2018)}]{kempf2018four}%
  \BibitemOpen
  \bibfield  {author} {\bibinfo {author} {\bibfnamefont {A.}~\bibnamefont
  {Kempf}},\ }\bibfield  {title} {\bibinfo {title} {Four aspects of
  superoscillations},\ }\href {https://doi.org/10.1007/s40509-018-0160-3}
  {\bibfield  {journal} {\bibinfo  {journal} {Quantum Studies: Mathematics and
  Foundations}\ }\textbf {\bibinfo {volume} {5}},\ \bibinfo {pages} {477}
  (\bibinfo {year} {2018})}\BibitemShut {NoStop}%
\bibitem [{\citenamefont {Ferreira}\ \emph {et~al.}(2007)\citenamefont
  {Ferreira}, \citenamefont {Kempf},\ and\ \citenamefont
  {Reis}}]{ferreira2007construction}%
  \BibitemOpen
  \bibfield  {author} {\bibinfo {author} {\bibfnamefont {P.}~\bibnamefont
  {Ferreira}}, \bibinfo {author} {\bibfnamefont {A.}~\bibnamefont {Kempf}},\
  and\ \bibinfo {author} {\bibfnamefont {M.}~\bibnamefont {Reis}},\ }\bibfield
  {title} {\bibinfo {title} {Construction of aharonov-berry's
  superoscillations},\ }\href {https://doi.org/10.1088/1751-8113/40/19/013}
  {\bibfield  {journal} {\bibinfo  {journal} {Journal of Physics A:
  Mathematical and Theoretical}\ }\textbf {\bibinfo {volume} {40}},\ \bibinfo
  {pages} {5141} (\bibinfo {year} {2007})}\BibitemShut {NoStop}%
\bibitem [{\citenamefont {Chojnacki}\ and\ \citenamefont
  {Kempf}(2016)}]{chojnacki2016new}%
  \BibitemOpen
  \bibfield  {author} {\bibinfo {author} {\bibfnamefont {L.}~\bibnamefont
  {Chojnacki}}\ and\ \bibinfo {author} {\bibfnamefont {A.}~\bibnamefont
  {Kempf}},\ }\bibfield  {title} {\bibinfo {title} {New methods for creating
  superoscillations},\ }\href {https://doi.org/10.1088/1751-8113/49/50/505203}
  {\bibfield  {journal} {\bibinfo  {journal} {Journal of Physics A:
  Mathematical and Theoretical}\ }\textbf {\bibinfo {volume} {49}},\ \bibinfo
  {pages} {505203} (\bibinfo {year} {2016})}\BibitemShut {NoStop}%
\bibitem [{\citenamefont {Šoda}\ and\ \citenamefont
  {Kempf}(2020)}]{soda_efficient_2020}%
  \BibitemOpen
  \bibfield  {author} {\bibinfo {author} {\bibfnamefont {B.}~\bibnamefont
  {Šoda}}\ and\ \bibinfo {author} {\bibfnamefont {A.}~\bibnamefont {Kempf}},\
  }\bibfield  {title} {\bibinfo {title} {Efficient method to create
  superoscillations with generic target behavior},\ }\href
  {https://doi.org/10.1007/s40509-020-00226-6} {\bibfield  {journal} {\bibinfo
  {journal} {Quantum Studies: Mathematics and Foundations}\ }\textbf {\bibinfo
  {volume} {7}},\ \bibinfo {pages} {347} (\bibinfo {year} {2020})}\BibitemShut
  {NoStop}%
\bibitem [{\citenamefont {Aharonov}\ \emph {et~al.}(2021)\citenamefont
  {Aharonov}, \citenamefont {Colombo}, \citenamefont {Sabadini}, \citenamefont
  {Shushi}, \citenamefont {Struppa},\ and\ \citenamefont
  {Tollaksen}}]{aharonov2021new}%
  \BibitemOpen
  \bibfield  {author} {\bibinfo {author} {\bibfnamefont {Y.}~\bibnamefont
  {Aharonov}}, \bibinfo {author} {\bibfnamefont {F.}~\bibnamefont {Colombo}},
  \bibinfo {author} {\bibfnamefont {I.}~\bibnamefont {Sabadini}}, \bibinfo
  {author} {\bibfnamefont {T.}~\bibnamefont {Shushi}}, \bibinfo {author}
  {\bibfnamefont {D.~C.}\ \bibnamefont {Struppa}},\ and\ \bibinfo {author}
  {\bibfnamefont {J.}~\bibnamefont {Tollaksen}},\ }\bibfield  {title} {\bibinfo
  {title} {A new method to generate superoscillating functions and
  supershifts},\ }\href {https://doi.org/10.1098/rspa.2021.0020} {\bibfield
  {journal} {\bibinfo  {journal} {Proceedings of the Royal Society A}\ }\textbf
  {\bibinfo {volume} {477}},\ \bibinfo {pages} {20210020} (\bibinfo {year}
  {2021})}\BibitemShut {NoStop}%
\bibitem [{\citenamefont {Tang}\ \emph {et~al.}(2016)\citenamefont {Tang},
  \citenamefont {Garg},\ and\ \citenamefont
  {Kempf}}]{tangScalingPropertiesSuperoscillations2016}%
  \BibitemOpen
  \bibfield  {author} {\bibinfo {author} {\bibfnamefont {E.}~\bibnamefont
  {Tang}}, \bibinfo {author} {\bibfnamefont {L.}~\bibnamefont {Garg}},\ and\
  \bibinfo {author} {\bibfnamefont {A.}~\bibnamefont {Kempf}},\ }\bibfield
  {title} {\bibinfo {title} {Scaling properties of superoscillations and the
  extension to periodic signals},\ }\href
  {https://doi.org/10.1088/1751-8113/49/33/335202} {\bibfield  {journal}
  {\bibinfo  {journal} {Journal of Physics A: Mathematical and Theoretical}\
  }\textbf {\bibinfo {volume} {49}},\ \bibinfo {pages} {335202} (\bibinfo
  {year} {2016})}\BibitemShut {NoStop}%
\bibitem [{\citenamefont {Chremmos}\ and\ \citenamefont
  {Fikioris}(2015)}]{chremmos_superoscillations_2015}%
  \BibitemOpen
  \bibfield  {author} {\bibinfo {author} {\bibfnamefont {I.}~\bibnamefont
  {Chremmos}}\ and\ \bibinfo {author} {\bibfnamefont {G.}~\bibnamefont
  {Fikioris}},\ }\bibfield  {title} {\bibinfo {title} {Superoscillations with
  arbitrary polynomial shape},\ }\href
  {https://doi.org/10.1088/1751-8113/48/26/265204} {\bibfield  {journal}
  {\bibinfo  {journal} {Journal of Physics A: Mathematical and Theoretical}\
  }\textbf {\bibinfo {volume} {48}},\ \bibinfo {pages} {265204} (\bibinfo
  {year} {2015})}\BibitemShut {NoStop}%
\bibitem [{\citenamefont {Ferreira}\ and\ \citenamefont
  {Kempf}(2006)}]{ferreiraSuperoscillationsFasterNyquist2006}%
  \BibitemOpen
  \bibfield  {author} {\bibinfo {author} {\bibfnamefont {P.}~\bibnamefont
  {Ferreira}}\ and\ \bibinfo {author} {\bibfnamefont {A.}~\bibnamefont
  {Kempf}},\ }\bibfield  {title} {\bibinfo {title} {Superoscillations: {{Faster
  Than}} the {{Nyquist Rate}}},\ }\href
  {https://doi.org/10.1109/TSP.2006.877642} {\bibfield  {journal} {\bibinfo
  {journal} {IEEE Transactions on Signal Processing}\ }\textbf {\bibinfo
  {volume} {54}},\ \bibinfo {pages} {3732} (\bibinfo {year}
  {2006})}\BibitemShut {NoStop}%
\bibitem [{\citenamefont {Jordan}\ \emph {et~al.}(2022)\citenamefont {Jordan},
  \citenamefont {Aharonov}, \citenamefont {Struppa}, \citenamefont {Colombo},
  \citenamefont {Sabadini}, \citenamefont {Shushi}, \citenamefont {Tollaksen},
  \citenamefont {Howell},\ and\ \citenamefont
  {Vamivakas}}]{jordanSuperphenomenaArbitraryQuantum2022}%
  \BibitemOpen
  \bibfield  {author} {\bibinfo {author} {\bibfnamefont {A.~N.}\ \bibnamefont
  {Jordan}}, \bibinfo {author} {\bibfnamefont {Y.}~\bibnamefont {Aharonov}},
  \bibinfo {author} {\bibfnamefont {D.~C.}\ \bibnamefont {Struppa}}, \bibinfo
  {author} {\bibfnamefont {F.}~\bibnamefont {Colombo}}, \bibinfo {author}
  {\bibfnamefont {I.}~\bibnamefont {Sabadini}}, \bibinfo {author}
  {\bibfnamefont {T.}~\bibnamefont {Shushi}}, \bibinfo {author} {\bibfnamefont
  {J.}~\bibnamefont {Tollaksen}}, \bibinfo {author} {\bibfnamefont {J.~C.}\
  \bibnamefont {Howell}},\ and\ \bibinfo {author} {\bibfnamefont {A.~N.}\
  \bibnamefont {Vamivakas}},\ }\href@noop {} {\bibinfo {title} {Super-phenomena
  in arbitrary quantum observables}} (\bibinfo {year} {2022}),\ \Eprint
  {https://arxiv.org/abs/2209.05650} {arXiv:2209.05650 [math-ph,
  physics:quant-ph]} \BibitemShut {NoStop}%
\bibitem [{\citenamefont {Cerjan}(2007)}]{Cerjan:07}%
  \BibitemOpen
  \bibfield  {author} {\bibinfo {author} {\bibfnamefont {C.}~\bibnamefont
  {Cerjan}},\ }\bibfield  {title} {\bibinfo {title} {Zernike-bessel
  representation and its application to hankel transforms},\ }\href
  {https://doi.org/10.1364/JOSAA.24.001609} {\bibfield  {journal} {\bibinfo
  {journal} {J. Opt. Soc. Am. A}\ }\textbf {\bibinfo {volume} {24}},\ \bibinfo
  {pages} {1609} (\bibinfo {year} {2007})}\BibitemShut {NoStop}%
\end{thebibliography}%

\end{document}